\documentclass[acmsmall,screen,nonacm]{acmart}
\setcopyright{none}
\renewcommand\footnotetextcopyrightpermission[1]{}

\usepackage{tabularx}
\usepackage{colortbl}
\usepackage{longtable}
\usepackage{array}
\usepackage{enumitem}
\usepackage{makecell}
\usepackage{wrapfig}
\usepackage{tikz}
\usepackage{tabularx}
\usetikzlibrary{shapes.geometric, arrows.meta, positioning, fit, backgrounds, calc}
\pgfdeclarelayer{midground}
\pgfsetlayers{background,midground,main}

\definecolor{stage1}{HTML}{2E5C8A}
\definecolor{stage2}{HTML}{3C7A4E}
\definecolor{stage3}{HTML}{B5582E}
\definecolor{purplestat}{HTML}{8268A8}
\definecolor{darktext}{HTML}{222222}
\definecolor{greytext}{HTML}{5A5A5A}
\definecolor{bordergrey}{HTML}{BFBFC4}
\definecolor{pillgrey}{HTML}{A8A8AC}

\usepackage{pgfplots}
\pgfplotsset{compat=1.17}   
\usepackage{xcolor}         

\newsavebox{\studyworkflowbox}

\newcommand{\qian}[1]{\textcolor{violet}{[Qian: #1]}}

\begin{document}

\title{Are We Testing the Wrong Quantum Bugs? A Study of Fault Representation Gaps in Quantum Software Testing}


\author{Asif Mohammed Saad}
\email{asaad2@tulane.edu}
\affiliation{%
  \institution{Tulane University}
  \city{New Orleans}
  \state{Louisiana}
  \country{USA}
}

\author{Ahsanul Ameen Sabit}
\email{AhsanulAmeen.Sabit@UTDallas.edu}
\affiliation{%
  \institution{The University of Texas at Dallas}
  \city{Richardson}
  \state{Texas}
  \country{USA}
}

\author{Jonah Brandenburgh}
\email{dal466132@utdallas.edu}
\affiliation{%
  \institution{The University of Texas at Dallas}
  \city{Richardson}
  \state{Texas}
  \country{USA}
}

\author{Jiyuan Wang}
\email{wjiyuan@tulane.edu}
\affiliation{%
  \institution{Tulane University}
  \city{New Orleans}
  \state{Louisiana}
  \country{USA}
}

\author{Qian Zhang}
\email{qzhang@cs.ucr.edu}
\affiliation{%
  \institution{University of California, Riverside}
  \city{Riverside}
  \state{California}
  \country{USA}
}

\author{Soneya Binta Hossain}
\email{sbhossain@utdallas.edu}
\affiliation{%
  \institution{The University of Texas at Dallas}
  \city{Richardson}
  \state{Texas}
  \country{USA}
}

\renewcommand{\shortauthors}{Author et al.}


\begin{abstract}
Quantum software relies on compilers, intermediate representations,
and execution frameworks that transform programs before execution.
Faults in these components can change the behavior of otherwise
correct quantum programs. Yet quantum testing techniques are often
evaluated using mutation-generated faults and curated bug benchmarks.
We investigate how well the faults represented by these artifacts
reflect those encountered in modern quantum software.

We audit four quantum mutation-testing tools and three curated
quantum bug benchmarks, and collect 410 verified bugs from ten
quantum software repositories over two years. We classify faults
from all three sources into a common taxonomy of 56 bug families
and compare their representation across mutation operators,
curated benchmarks, and recent real-world bugs.

The mutation tools support 20 bug families, and curated benchmarks
add 15 more. Yet only 56 of the 410 real-world bugs (13.7\%)
belong to families represented by either source; the remaining
354 (86.3\%) belong to families absent from both. The largest
unrepresented families concern compiler transformations (66 bugs),
type handling (53), and input validation (47). Unlike local gate
or statement changes, these faults often depend on intermediate
representations, compiler pipelines, input types, or dependency
versions.

Our findings reveal a substantial gap between the fault models
used to evaluate quantum testing techniques and the faults
observed in quantum programming systems. Existing mutation
operators primarily modify circuits and program constructs,
whereas many real-world faults arise in the compilers,
representations, and framework interfaces that process them.
Addressing this gap requires looking beyond circuit-level
faults toward testing semantic preservation across compiler
transformations, correctness at representation boundaries,
and interactions between software components. Our study
provides empirical evidence to guide the development of
fault models and benchmarks that capture these mechanisms
and preserve the contexts needed to reproduce them.
\end{abstract}




\maketitle

\section{Introduction}

Quantum programs depend on more than the circuits that developers write.
Before execution, quantum software passes through compilers, intermediate
representations, optimization passes, and execution frameworks that
transform and interpret the program~\cite{ibm_transpilation_docs,
cirq_basics_docs,microsoft_qsharp_docs}. A fault in any of these
components can change program behavior even when the original circuit
is correct. Bugs can also arise from how frameworks handle input types,
API contracts, and library dependencies~\cite{paltenghi2022bugs,
luo2022comprehensive,bensoussan2026taxonomy}. Testing quantum software
therefore involves not only checking the correctness of circuits, but
also the systems that construct, transform, and execute them.

Researchers commonly evaluate quantum testing techniques using two
kinds of fault artifacts. Mutation-testing tools generate faulty
versions of quantum programs by applying predefined changes to gates,
measurements, parameters, or program statements~\cite{
fortunato2022qmutpy,mendiluze2021muskit,garcia2024qcrmut}.
Curated bug benchmarks preserve faults collected from existing
quantum software projects~\cite{zhao2023bugs4q,bugs4qna_repo,
zhao2021bugpatterns,bugpattern_repo}. Both are important for
reproducible evaluation: mutants provide controlled faults, while
curated benchmarks provide examples of faults encountered in practice.
Techniques such as Quito~\cite{wang2021quito},
QuSBT~\cite{wang2022qusbt}, and MutTG~\cite{
wang2022mutationgeneration} have been evaluated using seeded faults,
and recent work has generated more than 700,000 mutated circuits
for testing studies~\cite{mendiluze2025mutants}. Bugs4Q similarly
provides reproducible real bugs for evaluating quantum testing and
debugging techniques~\cite{zhao2021bugs4q}. However, the conclusions
drawn from these evaluations depend on which faults the artifacts
represent~\cite{just2014mutants,jia2011mutation}.

Prior empirical studies have shown that quantum software faults
extend beyond individual circuit operations. Paltenghi and Pradel
studied 223 bugs from 18 quantum platforms and found quantum-specific
faults particularly common in components that represent, compile,
and optimize quantum abstractions~\cite{paltenghi2022bugs}.
Luo et al. examined 96 bug fixes across Qiskit, Cirq, Q\#, and
ProjectQ, more than 80\% of which involved quantum-specific
bugs~\cite{luo2022comprehensive}. Bensoussan et al. screened
more than 5,000 closed issues and classified 133 bugs into
47 subcategories~\cite{bensoussan2026taxonomy}. These studies
establish the diversity of faults in quantum software, but leave
open whether the artifacts used to evaluate quantum testing
techniques represent the same fault mechanisms.

There are reasons to expect a gap. Existing quantum mutation tools
primarily modify circuits or the programs that construct them; they
do not introduce faults into the implementations of the compilers
and frameworks processing those programs~\cite{
fortunato2022qmutpy,mendiluze2021muskit,garcia2024qcrmut,
shao2026quanforge}. Curated benchmarks include real faults, but
their contents reflect the projects and software versions from
which they were collected~\cite{zhao2023bugs4q,bugs4qna_repo,
zhao2021bugpatterns,bugpattern_repo,paltenghi2022bugs,
quantumplatformbugs_repo}. As frameworks evolve, new faults can
arise in compiler transformations, data representations, and
component interactions that these artifacts do not capture.
The question is not whether existing artifacts are useful, but
which fault mechanisms they represent and which they leave out.

This motivates our central research question:
\emph{How well do existing quantum mutation operators and curated
bug benchmarks represent the faults encountered in current quantum
software?} We investigate this question through three research
questions:

\begin{description}[style=multiline,leftmargin=3.4em,
labelwidth=2.8em,labelsep=0.6em]
    \item[\textbf{RQ1.}] What bug families are supported by the
    selected \textit{quantum mutation-testing tools}?

    \item[\textbf{RQ2.}] What bug families appear in
    well-established \textit{curated real quantum bug datasets}?

    \item[\textbf{RQ3.}] How do mutation-supported and curated
    bug families compare with recent \textit{in-the-wild quantum
    bugs}, and what lessons does this comparison provide for the
    community?
\end{description}

To answer these questions, we compare three sources of fault
evidence. First, we audit the mutation operators implemented by
four quantum testing tools: QMutPy~\cite{fortunato2022qmutpy},
Muskit~\cite{mendiluze2021muskit}, QCRMut~\cite{
garcia2024qcrmut}, and QuanForge~\cite{shao2026quanforge}.
Second, we examine three curated real-bug sources:
Bugs4Q-NA~\cite{bugs4qna_repo,zhao2023bugs4q},
Bug-Pattern~\cite{bugpattern_repo,zhao2021bugpatterns},
and Bugs in Quantum Computing Platforms~\cite{
quantumplatformbugs_repo,paltenghi2022bugs}.
Third, we collect bug reports filed between July 24, 2024,
and July 23, 2026, from ten repositories spanning different
parts of the quantum software stack. Two authors independently
reviewed the candidate reports, resolved disagreements through
consensus, and retained 410 verified bugs. We classify faults
from all three sources into a common taxonomy of 56 bug families,
allowing us to compare the fault mechanisms represented by
mutation operators, curated benchmarks, and recent repository bugs.

Our comparison reveals a substantial representation gap. The four
mutation tools support 20 bug families, and the curated benchmarks
add 15 more. Yet only 56 of the 410 recent bugs (13.7\%) belong
to families represented by either source. The remaining 354
(86.3\%) belong to families absent from both. The largest
unrepresented families are compiler transformation errors
(66 bugs), type-handling errors (53), missing input validation
(47), circuit-construction errors (29), and dependency
incompatibilities (28). These faults often depend on specific
compiler passes, intermediate representations, input types,
or library versions rather than an isolated gate or statement
change. 

The distinction has implications for how quantum programming
systems are tested and evaluated. Mutating a circuit can
reproduce a wrong gate or qubit target, but it does not recreate
a faulty compiler transformation or an incompatibility between
framework components. Some of these mechanisms may be modeled
through new operators that modify compiler logic or
representation handling. Others require regression cases
that preserve the triggering input, compiler configuration,
or dependency environment. Our findings therefore motivate
fault models and benchmarks that extend beyond individual
circuits to the programming systems responsible for their
execution.

The primary contributions of this paper are:

\begin{itemize}[leftmargin=*]
    \item 
    We examine four mutation-testing tools and identify
    20 supported bug families, including 11 quantum-specific
    and 9 inherited classical families.

    \item 
    We collect and independently review 410 bugs from ten
    quantum software repositories and classify faults from
    mutation tools, curated benchmarks, and recent repositories
    into 56 bug families.

    \item 
    We quantify the overlap between the three sources and
    identify 21 bug families absent from existing mutation
    tools and curated benchmarks, particularly in compiler
    transformations, input handling, program representations,
    and software dependencies.

    \item 
    We identify opportunities for testing compiler semantics,
    representation boundaries, and component interactions,
    and discuss which faults suit controlled mutation and which need context-preserving regression cases.
\end{itemize}

\section{Background \& Related Work}
\label{sec:related}

Quantum software testing research has produced empirical
bug studies, mutation-testing tools, and curated bug
benchmarks. These efforts provide different evidence about
software faults. Empirical studies characterize faults
observed in practice, mutation tools generate faults through
predefined transformations, and benchmarks preserve selected
real bugs. Our study examines the relationship between these
three sources of evidence. We first review the relevant work
and then explain what our comparison adds.

\subsection{The Quantum Software Stack and Its Bug Challenges}

Quantum software includes programming languages, circuit
libraries, compilers, simulators, and hardware-execution
services. Qiskit supports circuit construction and
compilation, Cirq provides circuit design and device-aware
execution, and Q\# integrates quantum operations with
classical control~\cite{ibm_transpilation_docs,
cirq_basics_docs,microsoft_qsharp_docs}. Correctness depends
not only on the circuit written by a developer, but also on
the representations and transformations used to process it.

Several studies have examined faults in these systems.
Paltenghi and Pradel~\cite{paltenghi2022bugs} studied
223 bugs from 18 quantum computing platforms and found
quantum-specific faults particularly common in components
that represent, compile, and optimize quantum abstractions.
They characterized the affected components, symptoms,
and recurring bug patterns. Our study builds on this
evidence: we include their published bug collection among
our curated sources, but ask a different question. Rather
than characterizing where faults occur, we determine
whether their underlying mechanisms are represented by
existing mutation operators and how those mechanisms
compare with bugs reported in recent repositories.

Luo et al.~\cite{luo2022comprehensive} analyzed 96 bug
fixes from Qiskit, Cirq, Q\#, and ProjectQ and found that
more than 80\% involved quantum-specific bugs.
Zhao et al.~\cite{zhao2023qmlbugs}, studying 391 bugs
from quantum machine-learning frameworks, reported a
much smaller quantum-specific proportion of 28\%.
Both studies characterize faults within their respective
subjects and demonstrate that the observed fault
distribution depends on the software being examined.
Our comparison instead focuses on fault mechanisms and
their representation across testing artifacts. We do not
compare the proportion of quantum-specific bugs alone;
we examine whether the mechanisms found in recent
software are present in mutation tools or curated datasets.

Bensoussan et al.~\cite{bensoussan2026taxonomy} provide
another closely related study. They analyzed more than
5,000 closed issues, identified 133 real faults, and
developed a taxonomy with seven main categories and
47 subcategories covering hybrid quantum-classical
architectures. Their taxonomy was further informed
and validated through developer interviews and expert
feedback. Their contribution is a classification of
faults and their characteristics in hybrid architectures.
Our contribution is different: we use a common
family-level classification to compare what mutation
operators can introduce, what curated benchmarks contain,
and what recent repository bugs exhibit. Thus, we
measure the representation of observed fault mechanisms
in existing testing artifacts, rather than developing
and validating a taxonomy alone.

Quetschlich and Di Matteo~\cite{
quetschlich2025classification} examined 14 quantum
bugs and found that many involved interactions between
multiple aspects of an algorithm or workflow. Their
study explains why individual bugs can be difficult
to classify and debug. We account for this possibility
by allowing a repository bug to belong to multiple
families when its evidence shows distinct mechanisms.
We then use these assignments to measure representation
across the three sources, rather than examining
debugging strategies for individual cases.

Other research directly tests or verifies quantum
compilers. QDiff~\cite{wang2021qdiff} and
MorphQ~\cite{paltenghi2023morphq} detect platform
faults by comparing executions of semantically equivalent
programs, while QASMInfer~\cite{lee2025exact}
provides exact circuit outcomes for testing.
Formal approaches verify circuit optimization
and compilation~\cite{hietala2021voqc,
tao2022giallar,li2022vqo} or synthesize
correctness-preserving optimizations~\cite{
xu2022quartz,xu2023queso}. These works develop
techniques for checking program or compiler
correctness. We do not propose another correctness
checker. Instead, we examine how frequently the
fault mechanisms encountered in programming systems
are represented by the artifacts used to evaluate
testing techniques.

\subsection{Quantum Mutation Testing}

Quantum mutation testing introduces controlled changes
to programs or circuits and checks whether a test suite
detects them~\cite{jia2011mutation}. Existing tools
differ in the artifacts and transformations they support.
Muskit~\cite{mendiluze2021muskit} introduces gate
addition, deletion, and replacement into circuits.
QCRMut~\cite{garcia2024qcrmut} generates random
circuit mutations while allowing selected positions
to remain unchanged. QMutPy~\cite{
fortunato2022qmutpy} mutates Python programs that
construct quantum circuits, combining quantum-specific
gate and measurement operators with classical operators
inherited from MutPy. QuanForge~\cite{
shao2026quanforge} targets trained quantum neural
networks by modifying gates and parameters after training.

These tools establish ways to generate faulty quantum
programs. Our study examines a different property of
their operators: which real-world fault mechanisms
their transformations can represent. We inspect the
operators of all four tools, map their transformations
to bug families, and compare those families with
verified repository bugs. This distinguishes the
number and variety of generated mutants from the
range of real fault mechanisms they model.

Prior mutation studies have primarily examined
the effectiveness or characteristics of generated
mutants. Fortunato et al.~\cite{
fortunato2022casestudy} studied how coverage and
assertions affect mutation scores for 24 Qiskit
programs. Wang et al.~\cite{
wang2022mutationgeneration} used mutants as
objectives for multi-objective test generation.
Mendiluze Usandizaga et al.~\cite{
mendiluze2025mutants} analyzed more than
700,000 mutants derived from 382 real-world
circuits, examining how operator choice, gate
type, mutation position, and circuit characteristics
relate to mutant survival. QMutBench~\cite{
mendiluze2026qmutbench} makes this collection
available as a reusable mutant benchmark.

These studies help researchers select and evaluate
mutants, including mutants with different detection
difficulties. They do not, however, establish how
closely the supported mutation mechanisms correspond
to faults reported in current quantum frameworks.
Our study addresses this question by comparing
operator-supported families with real bug families,
rather than measuring mutation scores or mutant
survival rates. This distinction also matters because
the selected operators primarily modify programs
and circuits, whereas many repository bugs occur
in the frameworks that process them.

\subsection{Curated Quantum Bug Benchmarks}

Curated benchmarks preserve real faults for
reproducible testing and debugging experiments.
Zhao et al.~\cite{zhao2021bugpatterns}
identified recurring Qiskit bug patterns,
including register mismatches, unrecognized
custom gates, repeated measurements, and
incorrect operations after measurement.
Bugs4Q~\cite{zhao2023bugs4q} contains
42 manually validated Qiskit bugs, including
30 bug--fix pairs with unit tests. These
artifacts make individual real faults available
for controlled evaluation, unlike synthetic
mutation-based datasets such as
QMutBench~\cite{mendiluze2026qmutbench}.

Our study examines three curated sources:
Bugs4Q-NA~\cite{bugs4qna_repo}, the
Bug-Pattern repository~\cite{bugpattern_repo},
and Bugs in Quantum Computing
Platforms~\cite{quantumplatformbugs_repo}.
They preserve different kinds of evidence,
including bug reports, code changes, tests,
and fault descriptions. We use these records
to identify which fault families the selected
curated sources represent.

The distinction from benchmark-construction
work is important. Bugs4Q and its extensions
aim to preserve useful real faults, while
Bug-Pattern documents recurring programming
mistakes. Our study does not introduce another
collection of isolated examples for the same
purpose. Instead, we treat these sources as
an existing representation of real faults and
ask which mechanisms they contain, which
they share with mutation tools, and which
mechanisms observed in recent repositories
are absent from both.

Recent work by Shi et al.~\cite{
shi2026qbugsset} introduces QBugsSet, a
lifecycle-based dataset of 2,429 bugs from
the Qiskit, Cirq, and PennyLane ecosystems.
They investigate how quantum-specific and
classical bugs differ in their lifecycle,
repair characteristics, and debugging
difficulty. Their study therefore provides
evidence about how bugs evolve and are
repaired. We investigate a different
dimension: whether the fault mechanisms
found in recent repositories are represented
by the mutation operators and curated
benchmarks selected for our comparison.
QBugsSet is not one of the three curated
sources in our baseline.

\subsection{Positioning of This Work}

The closest empirical studies already establish
that quantum faults occur throughout the
software stack. Paltenghi and Pradel identify
where platform bugs occur and characterize
their patterns; Bensoussan et al. classify
faults in hybrid quantum-classical
architectures; and Shi et al. examine the
lifecycle and repair of quantum bugs.
Mutation-testing studies, in contrast,
characterize the mutants that existing
operators can generate and their usefulness
for evaluating tests.

Our study connects these lines of research
through a comparison that none of them
directly performs. We map mutation
operators, curated real-bug cases, and
recent repository bugs into the same
56 bug families. This lets us determine
which families are supported by mutation
operators, which are represented only by
curated benchmarks, and which are absent
from both earlier sources. The comparison
is at the family level: membership in a
supported family does not establish that
an operator can reproduce every individual
bug in that family or that a test will
detect it.

Two examples illustrate why this comparison
matters. PennyLane issue \#9185
(\texttt{Q2Y0174}) reports that ZX circuit
transformations fail with PyZX 0.10.0
because of changes in qubit-label
registration~\cite{pennylane9185issue,
pennylane9179fix}. We classify it as
\emph{Dependency incompatibility}.
The failure arises from an interaction
between library versions, not an incorrect
gate in the input circuit.

Similarly, Qiskit issue \#15613
(\texttt{Q2Y0368}) reports that QPY
cannot reload a circuit containing a
register whose name begins with
\texttt{ancilla}~\cite{qiskit15613issue}.
We classify it as \emph{Serialization
error}. Here, the failure occurs when
a program representation is reconstructed,
rather than when a gate is executed.
Neither family is represented by the
selected mutation operators or curated
benchmarks.

These examples illustrate mechanisms
outside the local program transformations
supported by existing mutation tools.
Our empirical comparison measures how
often such mechanisms occur in the
selected recent repositories. It thereby
provides evidence for extending fault
models beyond circuit edits toward
compiler transformations, representation
handling, and interactions between
software components. It also identifies
where future benchmarks need to preserve
the execution context required to
reproduce real faults.

\input{figures/overview_soneya}

\section{Our Approach}
\label{sec:approach}
We investigate how well existing quantum mutation-testing tools and curated bug benchmarks represent bugs reported in recent quantum software development. We examine three sources: (1) mutation operators from four tools, (2) real bugs from three curated benchmarks, and (3) GitHub issues from ten quantum software repositories. To compare these sources, we group faults by their underlying causes into common \emph{bug families}.

Figure~\ref{fig:workflow} outlines our approach in three stages, numbered as in the figure. In Stage~1 (\emph{Build the baseline}), we derive 35 bug families from the operators of four mutation tools and the cases of three curated benchmarks (Sections~\ref{sec:data-collection} and~\ref{sec:baseline}). In Stage~2 (\emph{Collect and classify}), we collect 1,019 recent GitHub issues, sort them into tiers, and use an AI coding agent to confirm candidate bugs and propose a family for each, either from the baseline or a new one (Section~\ref{sec:classify-wild}). In Stage~3 (\emph{Review and compare}), two authors review the 412 candidates and retain 410 bugs, and we compare their families with the baseline (Section~\ref{sec:results}).

\subsection{Data Collection}
\label{sec:data-collection}

\vspace{.3em}
\noindent{\bf Mutation Tools.}
We checked four quantum mutation tools:
QMutPy~\cite{fortunato2022qmutpy},
Muskit~\cite{mendiluze2021muskit},
QCRMut~\cite{garcia2024qcrmut}, and
QuanForge~\cite{shao2026quanforge}. We chose them because each has a concrete list of mutation operators, together with a
paper, documentation, or code that lets us inspect the operators one by one. These tools target diverse artifacts: Muskit and QCRMut mutate quantum circuits, QMutPy mutates the Python code that builds them, and QuanForge
mutates trained quantum neural networks. QuanForge changes the gates and gate parameters of networks built with PennyLane and compares the predictions of the original and mutated networks on test inputs. It works only with its own built-in model structures, so we did not execute it. Instead, we identified its operators from its paper and source code.

As detailed in Table~\ref{tab:mutation-tool-statistics}, for each tool, we collect documentation describing its mutation operators and source code confirming their implementations. For the three tools we executed (QMutPy, Muskit, and QCRMut), we also collect the generated mutants to observe how the operators behave. For QCRMut, we identified 4 mutation operators from its documentation and checked their implementations in the source code. Because QCRMut is executable, we further collected 220 resulting mutants to verify the changes introduced by these operators.

\begin{table}[htbp]
\centering
\caption{Mutation-testing tools and evidence collected for
operator analysis. Each repository is pinned to a Git commit
(in parentheses) so the audited code can be
reproduced.}
\label{tab:mutation-tool-statistics}
\footnotesize
\setlength{\tabcolsep}{4pt}
\renewcommand{\arraystretch}{1.2}
\begin{tabularx}{\linewidth}{
    >{\raggedright\arraybackslash}p{3.2cm}
    >{\raggedright\arraybackslash}X
    c
    c
    c
}
\toprule
\textbf{Tool} &
\textbf{Repository (commit)} &
\textbf{Operators} &
\textbf{Executable} &
\textbf{Mutants} \\
\midrule

QMutPy~\cite{fortunato2022qmutpy,fortunato2022casestudy} &
{\texttt{danielfobooss/mutpy}}
    (\texttt{9c902b3}) &
32 & Yes & 54 \\

Muskit~\cite{mendiluze2021muskit} &
{\texttt{Simula-COMPLEX/muskit}}
    (\texttt{5e697b4}) &
3 & Yes & 34 \\

QCRMut~\cite{garcia2024qcrmut} &
{\texttt{sinugarc/QCRMutPaper}}
    (\texttt{cc9b132}) &
4 & Yes & 220 \\

QuanForge~\cite{shao2026quanforge} &
{\texttt{MinqiShao/QuanForge}}
    (\texttt{a282a32}) &
9 & No & -- \\

\midrule
\textbf{Total} & &
\textbf{48} & \textbf{3 of 4} & \textbf{308} \\

\bottomrule
\end{tabularx}
\end{table}

\begin{table}[!b]
\centering
\caption{Curated quantum bug benchmarks and the evidence
collected from each.}
\label{tab:curated-statistics}
\footnotesize
\setlength{\tabcolsep}{4pt}
\renewcommand{\arraystretch}{1.15}
\begin{tabularx}{\linewidth}{
    >{\raggedright\arraybackslash}p{3.6cm}
    c
    >{\raggedright\arraybackslash}p{2.8cm}
    >{\raggedright\arraybackslash}X
}
\toprule
\textbf{Benchmark} &
\textbf{Cases} &
\textbf{Original labels} &
\textbf{Evidence collected} \\
\midrule
Bugs4Q-NA~\cite{bugs4qna_repo} &
64 &
Free-text bug type per case &
Bug reports, buggy and fixed code, tests \\
Bug-Pattern~\cite{zhao2021bugpatterns,bugpattern_repo} &
23 &
8 bug patterns &
Pattern descriptions, code examples \\
Bugs in Quantum Computing
Platforms~\cite{paltenghi2022bugs,quantumplatformbugs_repo} &
223 &
12 platform components &
Bug records, fixes, component labels \\
\midrule
\textbf{Total} &
\textbf{310} & & \\
\bottomrule
\end{tabularx}
\vspace{2pt}
\end{table}

\vspace{.3em}
\noindent\textbf{Curated Bug Benchmarks.}
We collected real bug cases from three sources: Bugs4Q-NA~\cite{bugs4qna_repo}, Bug-Pattern~\cite{bugpattern_repo}, and Bugs in Quantum Computing Platforms~\cite{quantumplatformbugs_repo}. We selected these sources because they provide inspectable evidence of real bugs, including buggy and fixed code, issue reports, tests, and existing classifications. These three datasets are complementary to each other. Bugs4Q-NA provides reproducible,
case-level bugs from GitHub issues and Quantum Computing Stack Exchange questions~\cite{qcse_site}. Bug-Pattern documents recurring bug patterns in Qiskit programs, which its repository groups into eight types~\cite{bugpattern_repo}. Bugs in Quantum Computing Platforms contains 223 bugs
from 18 open-source platforms
and labels each bug with the platform component it occurs in (Figure~2 in~\cite{paltenghi2022bugs}). Table~\ref{tab:curated-statistics} summarizes the number of bug cases, original classifications, and evidence collected from each benchmark.

\vspace{.3em}
\noindent\textbf{In-the-Wild Issues.}
We selected ten repositories spanning quantum programming frameworks (Qiskit~\cite{qiskit_repo}, Cirq~\cite{cirq_repo}, PyQuil~\cite{pyquil_repo}, and PennyLane~\cite{pennylane_repo}), compilers and hybrid programming (TKET~\cite{tket_repo} and CUDA-Q~\cite{cudaq_repo}), cloud execution (Amazon Braket~\cite{braket_repo}), error mitigation (Mitiq~\cite{mitiq_repo}), simulation (QuTiP~\cite{qutip_repo}), and quantum program representation (OpenQASM~\cite{openqasm_repo}). Together, they cover every layer listed in Section~2.1 (languages, circuit libraries, compilers, simulators, and hardware-execution services) as well as error-mitigation middleware.

We collected issues created between July 24, 2024, and July 23, 2026. Each repository marks bug reports with its own label, so we searched each one for that label. For example, \texttt{bug} in Qiskit, TKET, CUDA-Q, Mitiq, and OpenQASM, \texttt{bug :bug:} in PennyLane and PyQuil, \texttt{kind/bug-report} in Cirq, and \texttt{BUG} in QuTiP. One of them, Amazon Braket, uses only the issue type and has no bug label. We included open and closed issues, excluded pull requests, and deduplicated the issues. As shown in Table~\ref{tab:issue-example}, for each issue, we collected its title, description, comments, labels, issue type, state, and dates, together with the pull requests and commits linked to it. For each linked pull request, we also recorded whether it was merged, whether it closed the issue, and whether it changed tests. The labels, state, and pull-request records show how strong the evidence of a fix is, and this sets the issue's tier (explained in Section~\ref{sec:classify-wild}). The title, description, and comments tell what went wrong, and the diff shows how the developers fixed it, so that the agent and authors can have clues to identify the bug family. The queries returned 1,056 records, comprising 1,019 unique issue candidates after removing 37 duplicate retrievals. Figure~\ref{fig:issue-statistics} shows their distribution by repository and retrieval route. 

{\setlength{\intextsep}{6pt plus 2pt}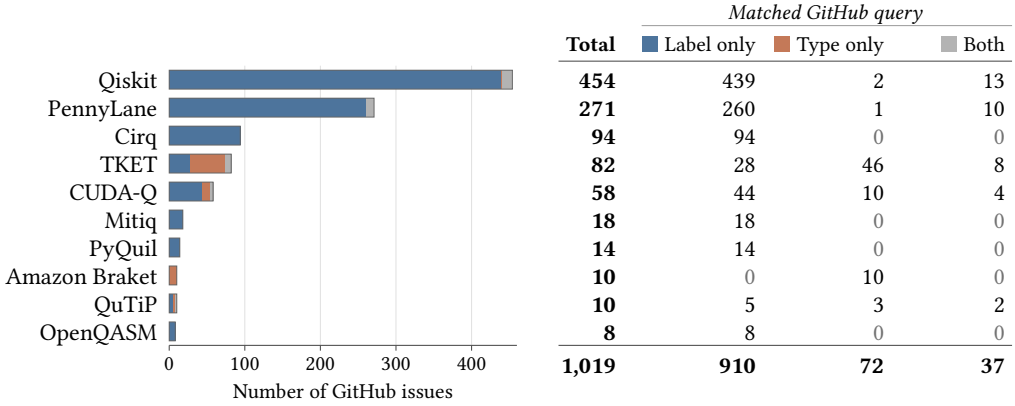
\begin{figure}[htbp]
\centering
\pgfmathsetlengthmacro{\ux}{4.6cm/460}%
\def\sw#1{{\color{#1}\rule[-0.15ex]{0.21cm}{0.21cm}}\hspace{2.5pt}}%
\begin{tikzpicture}[y=1cm, font=\small]
  \def\rowh{0.37}\def\barh{0.25}
  \def\xb{2.55}
  \def\cT{8.5}\def\cL{10.35}\def\cY{12.05}\def\cB{13.65}
  \def\bl{0.11}
  \pgfmathsetmacro{\ytop}{0.5*\rowh}
  \pgfmathsetmacro{\ybot}{-9*\rowh-0.5*\rowh}
  \node[anchor=base east, inner sep=1pt, font=\footnotesize\bfseries] at (\cT,\ytop+0.2) {Total};
  \node[anchor=base east, inner sep=1pt, font=\footnotesize] at (\cL,\ytop+0.2) {\sw{stage1!85}Label only};
  \node[anchor=base east, inner sep=1pt, font=\footnotesize] at (\cY,\ytop+0.2) {\sw{stage3!80}Type only};
  \node[anchor=base east, inner sep=1pt, font=\footnotesize] at (\cB,\ytop+0.2) {\sw{black!30}Both};
  \draw[black!70, line width=0.4pt] (\cT-0.8,\ytop+0.06) -- (\cB+0.05,\ytop+0.06);
  \node[anchor=base, inner sep=1pt, font=\footnotesize\itshape] at ({0.5*(\cL-1.55+\cB)},\ytop+0.62) {Matched GitHub query};
  \draw[black!40, line width=0.3pt] (\cL-1.55,\ytop+0.53) -- (\cB+0.05,\ytop+0.53);
  \foreach \v in {100,200,300,400}{\draw[black!12, line width=0.3pt] (\xb cm+\v*\ux,\ytop) -- (\xb cm+\v*\ux,\ybot);}
  \draw[black!70, line width=0.4pt] (\xb,\ybot) -- (\xb cm+460*\ux,\ybot);
  \foreach \v in {0,100,200,300,400}{
    \draw[black!70, line width=0.4pt] (\xb cm+\v*\ux,\ybot) -- ++(0,-0.07);
    \node[anchor=north, font=\footnotesize, inner sep=2pt] at (\xb cm+\v*\ux,\ybot-0.07) {\v};}
  \node[anchor=north, inner sep=2pt, font=\footnotesize] at (\xb cm+230*\ux,\ybot-0.42) {Number of GitHub issues};
  \foreach \name/\lab/\typ/\bo/\i in {Qiskit/439/2/13/0, PennyLane/260/1/10/1, Cirq/94/0/0/2, TKET/28/46/8/3, CUDA-Q/44/10/4/4, Mitiq/18/0/0/5, PyQuil/14/0/0/6, Amazon Braket/0/10/0/7, QuTiP/5/3/2/8, OpenQASM/8/0/0/9}{
    \pgfmathsetmacro{\yc}{-\i*\rowh}\pgfmathsetmacro{\ya}{\yc-0.5*\barh}\pgfmathsetmacro{\yb}{\yc+0.5*\barh}
    \pgfmathtruncatemacro{\tot}{\lab+\typ+\bo}
    \node[anchor=base east, inner sep=0pt] at (\xb-0.15,\yc-\bl) {\name};
    \ifnum\lab>0 \fill[stage1!85] (\xb,\ya) rectangle (\xb cm+\lab*\ux,\yb); \fi
    \ifnum\typ>0 \fill[stage3!80] (\xb cm+\lab*\ux,\ya) rectangle (\xb cm+\lab*\ux+\typ*\ux,\yb); \fi
    \ifnum\bo>0  \fill[black!30]  (\xb cm+\lab*\ux+\typ*\ux,\ya) rectangle (\xb cm+\tot*\ux,\yb); \fi
    \draw[black!60, line width=0.3pt] (\xb,\ya) rectangle (\xb cm+\tot*\ux,\yb);
    \node[anchor=base east, inner sep=1pt, font=\small\bfseries] at (\cT,\yc-\bl) {\tot};
    \ifnum\lab>0 \node[anchor=base east, inner sep=1pt] at (\cL,\yc-\bl) {\lab}; \else \node[anchor=base east, inner sep=1pt, text=black!55] at (\cL,\yc-\bl) {0}; \fi
    \ifnum\typ>0 \node[anchor=base east, inner sep=1pt] at (\cY,\yc-\bl) {\typ}; \else \node[anchor=base east, inner sep=1pt, text=black!55] at (\cY,\yc-\bl) {0}; \fi
    \ifnum\bo>0  \node[anchor=base east, inner sep=1pt] at (\cB,\yc-\bl) {\bo};  \else \node[anchor=base east, inner sep=1pt, text=black!55] at (\cB,\yc-\bl) {0}; \fi
  }
  \draw[black!70, line width=0.4pt] (\cT-0.8,\ybot) -- (\cB+0.05,\ybot);
  \node[anchor=base east, inner sep=1pt, font=\small\bfseries] at (\cT,\ybot-0.37) {1,019};
  \node[anchor=base east, inner sep=1pt, font=\small\bfseries] at (\cL,\ybot-0.37) {910};
  \node[anchor=base east, inner sep=1pt, font=\small\bfseries] at (\cY,\ybot-0.37) {72};
  \node[anchor=base east, inner sep=1pt, font=\small\bfseries] at (\cB,\ybot-0.37) {37};
\end{tikzpicture}
\caption{GitHub issues collected per repository, by the query that returned them: the repository's bug label, GitHub's \texttt{Bug} issue type, or both (counted once). Qiskit and PennyLane account for 725 of the 1,019 issues (71.1\%).}
\label{fig:issue-statistics}
\end{figure}

}

\begin{table}[tbp]
\centering
\caption{Running example, abbreviated. The upper part lists the fields our scripts collect for each issue, with their values for one issue and the steps that use them. The lower part lists what each step produced for this issue and who did it (Section~\ref{sec:classify-wild}).}
\label{tab:issue-example}
\footnotesize
\renewcommand{\arraystretch}{1.12}
\begin{tabularx}{\linewidth}{@{}>{\raggedright\arraybackslash\bfseries}p{2.3cm} >{\raggedright\arraybackslash}X >{\raggedright\arraybackslash\leavevmode\color{black!70}}p{2.6cm}@{}}
\toprule
\multicolumn{3}{@{}l}{\textit{Collected from GitHub by our scripts (Section~\ref{sec:data-collection})}} \\
\textit{\mdseries Field} & \textit{Value for this issue} & \textit{\color{black!70}Used for} \\
\midrule
\rowcolor{black!4}Issue title & Qiskit issue \#14726~\cite{qiskit14726issue}: ``Incorrect dimension in tensor product'' & Agent screening, family assignment \\
Label & \texttt{bug} & Finding bug reports and tier assignment \\
\rowcolor{black!4}Issue type & None & Finding bug reports \\
State & Closed as completed & Tier assignment \\
\rowcolor{black!4}Dates & Opened July 12, 2025. Closed November 10, 2025. & Keeping issues within the study period \\
Description (abbreviated) & ``The code below should return state \texttt{|1>} and it returns 4-dimensional vector instead \ldots''\par\vspace{2pt}
\colorbox{black!5}{\begin{minipage}{\dimexpr\linewidth-2\fboxsep\relax}\ttfamily\scriptsize
X = Operator(XGate())\\
sv0 = Statevector.from\_label('0')\\
\textcolor{black!60}{\# Apply X gate: X|0$\rangle$ = |1$\rangle$}\\
new\_state = X @ sv0
\end{minipage}}\par\vspace{1pt}\ldots & Agent screening \\
\rowcolor{black!4}Comments (3, abbreviated) & ``Seems like \texttt{Operator} doesn't have specialised code to apply itself to a \texttt{Statevector} \ldots{} it's just an oversight.'' & Agent screening \\
Linked pull request & \#14963~\cite{qiskit14963pr}: merged, closes the issue, from the same repository, changes three files including a test & Tier assignment, agent screening \\
\midrule
\multicolumn{3}{@{}l}{\textit{Produced for this issue by the pipeline (Section~\ref{sec:classify-wild})}} \\
\textit{\mdseries Step} & \textit{Output} & \textit{\color{black!70}Done by} \\
\midrule
\rowcolor{black!4}Tier assignment & Tier~1 & Tiering rule \\
Agent screening & Confirmed bug. Fix: \#14963. Quoted line: \texttt{return other.evolve(self)} & Agent \\
\rowcolor{black!4}Fix explanation & ``Operator matrix multiplication now dispatches \texttt{Statevector} operands through \texttt{evolve}, preserving the expected state dimensions and type.'' & Agent \\
Evidence check & Passed & Exact-match check \\
\rowcolor{black!4}Fault description & Location: API interface. Root cause: wrong target. Symptom: wrong output. Repair: retarget. & Agent \\
Family assignment & \emph{Type-handling error} & Agent \\
\rowcolor{black!4}Author review & Family kept & Both reviewers \\
\bottomrule
\end{tabularx}
\end{table}

\subsection{Constructing the Bug-Family Baseline}
\label{sec:baseline}
To compare bugs across sources, we group them by their underlying causes into a common set of \textbf{\em bug families}. We construct the bug-family for mutation operators and curated benchmarks in two steps, described below. The result is the \emph{baseline}: \textit{a fixed catalogue of 35 bug families (20 from mutation tools, 15 from curated benchmarks), each with a name and a definition of the mechanism change it
represents, built only from the mutation tools and curated benchmarks, fixed before we classified any repository bug, and used as the reference for classifying in-the-wild bugs.}

\vspace{.3em}
\noindent{\bf Step 1: Mapping Mutation Operators.}
We first map mutation operators to the faults introduced by their transformations. For example, Muskit's gate addition, deletion, and replacement operators correspond to \emph{Extra gate}, \emph{Missing gate}, and \emph{Wrong gate}, respectively. When an operator has multiple effects, we classify it by the primary fault it introduces. Thus, we classify QMutPy's \texttt{QuantumGateReplacement} as \emph{Wrong gate}, although it can also introduce \emph{Extra gate} faults. For example, in a circuit that prepares a Bell state, \texttt{QuantumGateReplacement} replaces the Hadamard gate with an X gate:
\begin{center}
\colorbox{black!5}{\begin{minipage}{0.9\linewidth}\ttfamily\footnotesize
qc = QuantumCircuit(2, 2)\\
qc.h(0)\hspace{8ex}\textcolor{black!60}{\# mutant: qc.x(0)}\\
qc.cx(0, 1)\\
qc.measure([0, 1], [0, 1])
\end{minipage}}
\end{center}
\noindent The mutant always measures \texttt{11}, whereas the original measures \texttt{00} or \texttt{11} with equal probability, so the mutant represents a \emph{Wrong gate} fault. As shown in Table~\ref{tab:baseline-mutation}, this process identifies 20 mutation-supported families: 11 quantum-specific and 9 classical families inherited by QMutPy from MutPy.

\begin{table}[htbp]
\centering
\caption{Bug families supported by four quantum mutation-testing
tools. A checkmark indicates that at least one mutation operator
supports the family.
}
\label{tab:baseline-mutation}
\footnotesize
\setlength{\tabcolsep}{3pt}
\renewcommand{\arraystretch}{1.05}

\begin{tabularx}{\linewidth}{
    >{\raggedright\arraybackslash}p{3.3cm}
    >{\raggedright\arraybackslash}X
    cccc
}
\toprule
\textbf{Bug Family} &
\textbf{Definition} &
\textbf{QMutPy} &
\textbf{Muskit} &
\textbf{QCRMut} &
\textbf{QuanForge} \\
\midrule

\multicolumn{6}{l}{\textit{Quantum-specific (11)}} \\
\addlinespace[2pt]

Extra gate &
Unnecessary gate insertion &
$\checkmark$ & $\checkmark$ & $\checkmark$ & $\checkmark$ \\

Extra measurement &
Unnecessary measurement insertion & $\checkmark$ & -- & -- & -- \\

Flipped parameter sign &
Reversal of a gate parameter's sign &
-- & -- & -- & $\checkmark$ \\

Misplaced gate &
Gate placed at an incorrect position &
-- & -- & -- & $\checkmark$ \\

Missing gate &
Removal of a required gate &
$\checkmark$ & $\checkmark$ & $\checkmark$ & $\checkmark$ \\

Missing measurement &
Removal of a required measurement &
$\checkmark$ & -- & -- & -- \\

Perturbed parameter &
Incorrect gate parameter value &
-- & $\checkmark$ & $\checkmark$ & $\checkmark$ \\

Swapped parameters &
Exchange of gate parameter positions &
-- & -- & -- & $\checkmark$ \\

Wrong gate &
Replacement with an incorrect gate &
$\checkmark$ & $\checkmark$ & $\checkmark$ & $\checkmark$ \\

Wrong gate arity &
Incorrect number of target qubits &
-- & -- & -- & $\checkmark$ \\

Wrong qubit target &
Operation applied to an incorrect qubit &
-- & -- & $\checkmark$ & $\checkmark$ \\

\midrule
\multicolumn{6}{l}{\textit{Classical (9)}} \\
\addlinespace[2pt]

Broken inheritance &
Incorrect parent-class method usage &
$\checkmark$ & -- & -- & -- \\

Mishandled exception &
Incorrect exception handling &
$\checkmark$ & -- & -- & -- \\

Missing statement &
Removal of a required statement &
$\checkmark$ & -- & -- & -- \\

Wrong arithmetic &
Incorrect arithmetic operator &
$\checkmark$ & -- & -- & -- \\

Wrong comparison &
Incorrect comparison operator &
$\checkmark$ & -- & -- & -- \\

Wrong conditional &
Incorrect condition negation &
$\checkmark$ & -- & -- & -- \\

Wrong constant &
Incorrect literal value &
$\checkmark$ & -- & -- & -- \\

Wrong decorator &
Incorrect or missing decorator &
$\checkmark$ & -- & -- & -- \\

Wrong loop control &
Incorrect loop iteration or termination &
$\checkmark$ & -- & -- & -- \\

\midrule
\multicolumn{6}{l}{\textit{Named in QuanForge's code but not implemented, so not counted (1)}} \\
\addlinespace[2pt]

Zeroed parameter &
Gate parameter set to zero &
-- & -- & -- & -- \\

\bottomrule
\end{tabularx}

\vspace{2pt}
\raggedright
\scriptsize
\end{table}

\vspace{.3em}
\noindent{\bf Step 2: Mapping Curated Bugs.}
We then map curated bug cases to the identified families based on their documented causes. We reuse an existing family when its definition matches the bug; otherwise, we introduce a new family. For example, an omitted gate maps to \emph{Missing gate}, whereas an omitted API initialization call introduces \emph{Missing method call}. We distinguish families by their underlying causes, even when their names are similar. The curated benchmarks cover all 20 mutation-supported families and add 15 more: \emph{Zeroed parameter}, which QuanForge names but does not implement, and the 14 families listed in Table~\ref{tab:curated-families}. Together they form a 35-family baseline for classifying in-the-wild bugs.

\begin{table}[tbp]
\centering
\caption{Fourteen of the 15 bug families that curated bug
benchmarks add beyond the mutation tools. The fifteenth,
\emph{Zeroed parameter}, is listed in Table~\ref{tab:baseline-mutation}. \emph{Cases} counts the curated cases assigned
to each family. A checkmark indicates that a benchmark
contains at least one case from the family.
B4Q: Bugs4Q-NA; BP: Bug-Pattern;
PCB: Bugs in Quantum Computing Platforms.}
\label{tab:curated-families}
\footnotesize
\setlength{\tabcolsep}{3pt}
\renewcommand{\arraystretch}{1.05}

\begin{tabularx}{\linewidth}{
    >{\raggedright\arraybackslash}p{3.8cm}
    >{\raggedright\arraybackslash}X
    c
    ccc
}
\toprule
\textbf{Bug Family} &
\textbf{Definition} &
\textbf{Cases} &
\textbf{B4Q} &
\textbf{BP} &
\textbf{PCB} \\
\midrule

API issue &
Incorrect API signature or behavior &
33 & $\checkmark$ & $\checkmark$ & $\checkmark$ \\

Conditional error &
Incorrect branch condition or missing case &
33 & $\checkmark$ & -- & $\checkmark$ \\

Confusing mutator with accessor &
Incorrect use of state-modifying or read-only methods &
8 & $\checkmark$ & -- & $\checkmark$ \\

Constant replacement &
Incorrect named constant or configuration value &
42 & $\checkmark$ & -- & $\checkmark$ \\

Incorrect operator &
Incorrect operator, predicate, or operation &
48 & $\checkmark$ & $\checkmark$ & $\checkmark$ \\

Incorrect parameters &
Incorrect argument value or default &
33 & $\checkmark$ & $\checkmark$ & $\checkmark$ \\

Missing decomposition &
Missing decomposition into supported gates &
5 & $\checkmark$ & -- & $\checkmark$ \\

Missing field set &
Missing or unpropagated object attribute &
10 & $\checkmark$ & -- & $\checkmark$ \\

Missing import &
Missing required module import &
6 & $\checkmark$ & $\checkmark$ & $\checkmark$ \\

Missing method call &
Omitted required API or initialization call &
8 & $\checkmark$ & $\checkmark$ & $\checkmark$ \\

Missing transpilation &
Missing compilation for target hardware &
4 & $\checkmark$ & -- & -- \\

Refactor &
Incomplete refactoring of internal references &
4 & $\checkmark$ & -- & $\checkmark$ \\

Registration error &
Missing registration of a custom operation &
1 & $\checkmark$ & $\checkmark$ & $\checkmark$ \\

Reversed output convention &
Incorrect result ordering convention &
4 & $\checkmark$ & -- & $\checkmark$ \\

\midrule
\textbf{Total} & &
\textbf{239}\, & \textbf{14}\, & \textbf{6}\, & \textbf{13}\, \\

\bottomrule
\end{tabularx}
\end{table}

\subsection{Screening and Classifying In-the-Wild Bug Issues}
\label{sec:classify-wild}

\vspace{.3em}
\noindent\textbf{Issue Candidate Screening.}
We partition the 1,019 issues into three mutually exclusive tiers based on their linked pull requests, their status, and their labels. Only bug labels such as \texttt{bug} and labels that decline an issue such as \texttt{wontfix} count, and topic labels are ignored. The tiers are ordered by decreasing confidence that an issue reports a bug:
\begin{itemize}
    \item \textbf{Tier~1 (473 issues):} Issues that a merged pull request from the same repository closed, that were not declined (for example, closed as not planned or labeled \texttt{wontfix}), and that have a bug label or a fix that changes a test. This is the strongest evidence that an issue reports a bug.
    \item \textbf{Tier~2 (133 issues):} Issues not in Tier~1 that meet at
    least one of two conditions: (i) they are linked to a merged pull request
    that did not close them, or (ii) they have been accepted by maintainers, but are not linked to any merged pull request. Either condition shows that maintainers recognized the issue.
    \item \textbf{Tier~3 (413 issues):} The remaining issues, which have not been accepted by maintainers. They are the least likely to report bugs, so we exclude them from further analysis.
\end{itemize}
The issue in Table~\ref{tab:issue-example} is in Tier~1, which makes it a strong bug candidate for the agent to check.

We built the curation pipeline and its custom agent with GitHub Copilot and Claude Opus 4.8, and the agent's runs used GPT-5-family models in OpenAI Codex. In \emph{agent screening}, the agent decides whether a Tier~1--2 issue reports a real bug that a merged pull request fixed. For each issue, it received the collected record (title, description, comments, labels, state, and linked pull requests) and was told to read the whole issue and the full diff and changed tests of each linked pull request. It had to choose one merged fix, preferring the original over a backport, and its instructions state that a title or a label alone does not prove a fix. The agent screened the 606 Tier~1--2 issues, and confirmed 529 as bugs, rejected 64, and left 13 undecided. For each confirmed bug, it recorded the fixing pull request, a quoted line from the fix, and a one-sentence explanation of how the change repairs the reported behavior. For Qiskit issue \#14726~\cite{qiskit14726issue} in Table~\ref{tab:issue-example}, it selected pull request \#14963~\cite{qiskit14963pr}, the only linked fix, and gave the changed line \texttt{return other.evolve(self)} as its \emph{quoted line}. Its explanation says that the fix routes \texttt{Statevector} operands through \texttt{evolve} (the \emph{Agent screening} and \emph{Fix explanation} rows). 

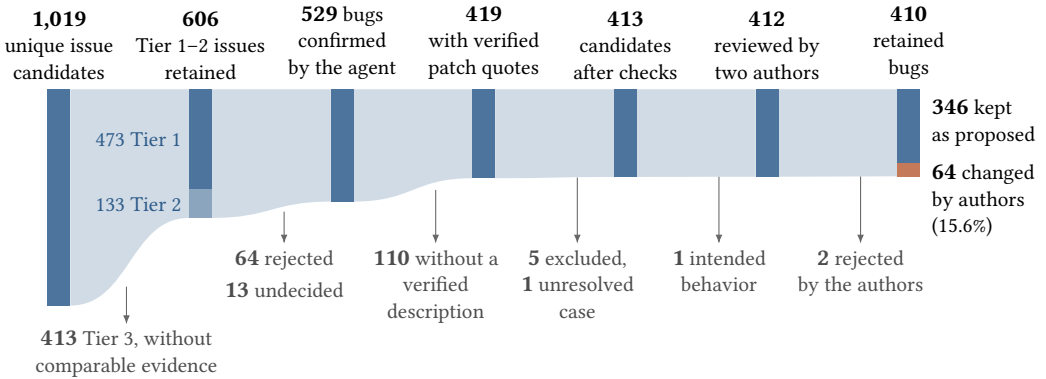
\begin{figure}[h]
\centering
\def\cnt#1{{\small\bfseries #1}}%
\def\dsc#1{{\footnotesize #1}}%
\pgfmathsetlengthmacro{\barw}{0.30cm}%
\pgfmathsetlengthmacro{\lmar}{0.75cm}
\pgfmathsetlengthmacro{\rmar}{1.62cm}
\pgfmathsetlengthmacro{\dx}{(\linewidth-\lmar-\rmar-\barw)/6}%
\pgfmathsetlengthmacro{\xA}{\lmar}%
\pgfmathsetlengthmacro{\xB}{\xA+\dx}%
\pgfmathsetlengthmacro{\xC}{\xA+2*\dx}%
\pgfmathsetlengthmacro{\xD}{\xA+3*\dx}%
\pgfmathsetlengthmacro{\xE}{\xA+4*\dx}%
\pgfmathsetlengthmacro{\xF}{\xA+5*\dx}%
\pgfmathsetlengthmacro{\xG}{\xA+6*\dx}%
\begin{tikzpicture}[x=1cm, y=1cm, font=\footnotesize,
  exit/.style={black!60, line width=0.4pt, -{Latex[length=1.2mm]}},
  exitlabel/.style={anchor=north, align=center, inner sep=2pt, text=black!70},
  station/.style={anchor=south, align=center, inner sep=2pt}]
  \def\s{0.0028}   
  \pgfmathsetmacro{\hA}{1019*\s}\pgfmathsetmacro{\hB}{606*\s}\pgfmathsetmacro{\hBa}{473*\s}
  \pgfmathsetmacro{\hC}{529*\s}\pgfmathsetmacro{\hD}{419*\s}\pgfmathsetmacro{\hE}{413*\s}
  \pgfmathsetmacro{\hF}{412*\s}\pgfmathsetmacro{\hG}{410*\s}\pgfmathsetmacro{\hGa}{346*\s}
  \pgfmathsetmacro{\ybase}{-(\hB+\hC)/2-0.45}   
  \newcommand{\band}[5]{\pgfmathsetlengthmacro{\xm}{(#2+#3)/2}%
    \fill[#1] (#2,0) -- (#3,0) -- (#3,-#5) .. controls (\xm,-#5) and (\xm,-#4) .. (#2,-#4) -- cycle;}
  \newcommand{\exitarrow}[7]{%
    \pgfmathsetlengthmacro{\ex}{#1+(#2-#1)*(1.5*#5-1.5*#5*#5+#5*#5*#5)}%
    \pgfmathsetmacro{\ey}{-(#3+(#4-#3)*(3*#5*#5-2*#5*#5*#5))}%
    \pgfmathsetmacro{\ty}{#6}%
    \edef\tmp{\noexpand\draw[exit] (\ex,\ey) -- (\ex,\ty)}%
    \tmp node[exitlabel] {#7};}
  \band{stage1!22}{\xA+\barw}{\xB}{\hA}{\hB}
  \band{stage1!22}{\xB+\barw}{\xC}{\hB}{\hC}
  \band{stage1!22}{\xC+\barw}{\xD}{\hC}{\hD}
  \band{stage1!22}{\xD+\barw}{\xE}{\hD}{\hE}
  \band{stage1!22}{\xE+\barw}{\xF}{\hE}{\hF}
  \band{stage1!22}{\xF+\barw}{\xG}{\hF}{\hG}
  \fill[stage1!85] (\xA,0) rectangle (\xA+\barw,-\hA);
  \fill[stage1!85] (\xB,0) rectangle (\xB+\barw,-\hBa);
  \fill[stage1!55] (\xB,-\hBa) rectangle (\xB+\barw,-\hB);
  \fill[stage1!85] (\xC,0) rectangle (\xC+\barw,-\hC);
  \fill[stage1!85] (\xD,0) rectangle (\xD+\barw,-\hD);
  \fill[stage1!85] (\xE,0) rectangle (\xE+\barw,-\hE);
  \fill[stage1!85] (\xF,0) rectangle (\xF+\barw,-\hF);
  \fill[stage1!85] (\xG,0) rectangle (\xG+\barw,-\hGa);
  \fill[stage3!80] (\xG,-\hGa) rectangle (\xG+\barw,-\hG);
  \node[station] at (\xA+0.5*\barw,0.06) {\cnt{1,019}\\\dsc{unique issue}\\\dsc{candidates}};
  \node[station] at (\xB+0.5*\barw,0.06) {\cnt{606}\\\dsc{Tier 1--2 issues}\\\dsc{retained}};
  \node[station] at (\xC+0.5*\barw,0.06) {\cnt{529} \dsc{bugs}\\\dsc{confirmed}\\\dsc{by the agent}};
  \node[station] at (\xD+0.5*\barw,0.06) {\cnt{419}\\\dsc{with verified}\\\dsc{patch quotes}};
  \node[station] at (\xE+0.5*\barw,0.06) {\cnt{413}\\\dsc{candidates}\\\dsc{after checks}};
  \node[station] at (\xF+0.5*\barw,0.06) {\cnt{412}\\\dsc{reviewed by}\\\dsc{two authors}};
  \node[station] at (\xG+0.5*\barw,0.06) {\cnt{410}\\\dsc{retained}\\\dsc{bugs}};
  \node[anchor=east, inner sep=1pt, text=stage1] at (\xB-0.06cm,-0.5*\hBa) {473 Tier 1};
  \node[anchor=east, inner sep=1pt, text=stage1] at (\xB-0.06cm,{-0.5*(\hBa+\hB)}) {133 Tier 2};
  \node[anchor=west, align=left, inner sep=2pt] at (\xG+\barw+0.08cm,-0.5*\hGa+0.05) {\cnt{346} \dsc{kept}\\\dsc{as proposed}};
  \node[anchor=north west, align=left, inner sep=2pt] at (\xG+\barw+0.08cm,-\hGa+0.06) {\cnt{64} \dsc{changed}\\\dsc{by authors}\\\dsc{(15.6\%)}};
  \exitarrow{\xA+\barw}{\xB}{\hA}{\hB}{0.3}{\ey-0.5}{\cnt{413} Tier 3, without\\comparable evidence}
  \exitarrow{\xB+\barw}{\xC}{\hB}{\hC}{0.42}{\ybase}{\cnt{64} rejected\\[2pt]\cnt{13} undecided}
  \exitarrow{\xC+\barw}{\xD}{\hC}{\hD}{0.5}{\ybase}{\cnt{110} without a\\verified\\description}
  \exitarrow{\xD+\barw}{\xE}{\hD}{\hE}{0.5}{\ybase}{\cnt{5} excluded,\\\cnt{1} unresolved\\case}
  \exitarrow{\xE+\barw}{\xF}{\hE}{\hF}{0.5}{\ybase}{\cnt{1} intended\\behavior}
  \exitarrow{\xF+\barw}{\xG}{\hF}{\hG}{0.5}{\ybase}{\cnt{2} rejected\\by the authors}
\end{tikzpicture}
\caption{Screening and review pipeline (Section~\ref{sec:classify-wild}).
Bar heights are proportional to the number of issues at each step. The blue
band carries the retained issues, and gray arrows mark those that leave.
Rejected issues had no merged code fix, mostly because the linked change
touched only documentation, tests, or CI. Undecided issues had a fix in
another repository or one that could not be tied to the issue. Of the 1,019
unique issue candidates, 412 reach independent author review and 410 are
retained. The authors kept the proposed family of 346 bugs and changed it
for 64 (15.6\%).}
\label{fig:review-pipeline}
\end{figure}

Next, the agent described the confirmed bugs with four labels: the code location, the root cause, the symptom, and the kind of repair (the \emph{Fault description} row of Table~\ref{tab:issue-example}). The word ``transpile'', for instance, sets the location to the compiler. We set aside the remaining 110 bugs because their evidence did not show what and where the fault was, as when the quoted line only added an import or no diff could be retrieved. For each of the 419 described bugs, the quoted line occurs verbatim in the fixing diff (the \emph{Evidence check} row of Table~\ref{tab:issue-example}). For the issue in Table~\ref{tab:issue-example}, the \emph{quoted line} is found in the diff of pull request \#14963~\cite{qiskit14963pr}, which confirms that the evidence comes from the actual fix. The fault descriptions served only to propose families, and the two authors later checked every family against the issue and its fix. Our script then flagged 12 of the 419 bugs whose fixes looked non-behavioral. Two separate agent reviews excluded five of them, whose fixes changed only documentation or package metadata, and held one on which they disagreed. Finally, we excluded one report that describes intended behavior, leaving 412 candidates. Figure~\ref{fig:review-pipeline} summarizes the screening process.

\vspace{.3em}
\noindent\textbf{AI-Based Family Assignment.}
Before author review, \textit{gemini-3.1-flash-lite}~\cite{google2026gemini} proposed one or
more bug families for each candidate. A bug could receive more than one
family when its failure involved distinct mechanisms. The model received
the final 35-family baseline, including the name and definition of each
family, and five additional family definitions derived from our
observation of a random sample of repository issues:
\emph{Compiler transformation error}, \emph{Type-handling error},
\emph{Missing input validation}, \emph{Circuit construction error}, and
\emph{Dependency incompatibility}.

For each candidate, the model read the whole issue, including its
description, reproducer, error output, and discussion, together with
the linked fix, its diff, and its tests. It matched the reported fault
against the supplied family definitions rather than keywords. When no
supplied definition fit, it proposed a new family with inclusion and
exclusion criteria, reused that family for later similar bugs, and
quoted supporting evidence with a confidence level. The five additional families name faults that the baseline lacks, but they could also attract bugs that a baseline family fits. Because they were derived from our repository observations, the two authors independently checked every proposed family against the issue and its fix.

For Qiskit issue \#14726 (Table~\ref{tab:issue-example}), the model
proposed \emph{Type-handling error}, which covers code that assumes one
kind of value and fails on another valid one: the \texttt{@} operator
handled a \texttt{Statevector} as if it were another
\texttt{Operator}. The authors retained this family during review.

We also classified each bug  by the concerns needed to explain its fault. The AI coding agent classified each bug as
\emph{quantum-specific}, \emph{quantum--classical}, or
\emph{purely classical} based on whether the failure or repair
depends on quantum circuits, states, measurements, compilation,
or quantum-program representations. It is \emph{quantum--classical} when the fault needs both a quantum concern and a general software mechanism, such as type handling, input validation, or a dependency interaction. A quantum operation that fails only under a particular numerical library is one example. It is
\emph{purely classical} when neither its failure nor its repair depends on quantum semantics. For example, \textit{Wrong qubit target} is quantum-specific in 8 of its 10 bugs, \textit{Symbolic algebra error} is quantum--classical in 3 of its 4 bugs, and no family is mostly purely classical. The split for each family is in Figure~\ref{fig:family-specificity}.

\vspace{.3em}
\noindent\textbf{Independent Validation.}
Two authors independently reviewed all 412 candidates against the repository evidence and proposed labels. Each could revise the eligibility decision or add, remove, or rename families. We retained reports supported by a reproducer, stack trace, linked fix, or developer confirmation. We excluded feature requests, questions, duplicates, and reports lacking sufficient evidence. Disagreements were resolved by consensus; new families required agreement from both authors. 

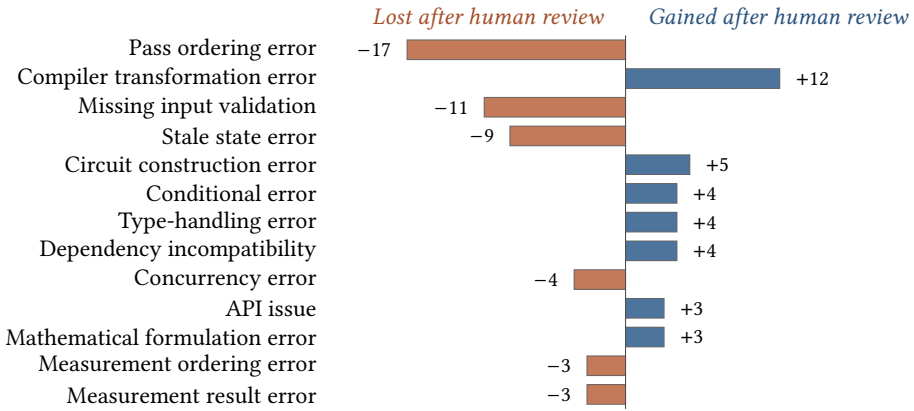
\begin{figure}[h]
\centering
\begin{tikzpicture}[font=\small, x=0.17cm, y=1cm]
  \def\rowh{0.38}
  \def\barh{0.26}
  \foreach \name/\delta/\i in {%
    Pass ordering error/-17/0,
    Compiler transformation error/12/1,
    Missing input validation/-11/2,
    Stale state error/-9/3,
    Circuit construction error/5/4,
    Conditional error/4/5,
    Type-handling error/4/6,
    Dependency incompatibility/4/7,
    Concurrency error/-4/8,
    API issue/3/9,
    Mathematical formulation error/3/10,
    Measurement ordering error/-3/11,
    Measurement result error/-3/12%
  }{
    \pgfmathsetmacro{\yc}{-\i*\rowh}
    \pgfmathsetmacro{\ya}{\yc-0.5*\barh}
    \pgfmathsetmacro{\yb}{\yc+0.5*\barh}
    \node[anchor=east, inner sep=0pt] at (-24,\yc) {\name};
    \ifnum\delta<0
      \fill[stage3!80] (\delta,\ya) rectangle (0,\yb);
      \draw[black!60, line width=0.3pt] (\delta,\ya) rectangle (0,\yb);
      \node[anchor=east, inner sep=0pt, font=\footnotesize] at (\delta-1.2,\yc) {$-$\number\numexpr-\delta\relax};
    \else
      \fill[stage1!85] (0,\ya) rectangle (\delta,\yb);
      \draw[black!60, line width=0.3pt] (0,\ya) rectangle (\delta,\yb);
      \node[anchor=west, inner sep=0pt, font=\footnotesize] at (\delta+1.2,\yc) {$+$\delta};
    \fi
  }
  \draw[black!70, line width=0.4pt] (0,0.5*\rowh) -- (0,{-12*\rowh-0.5*\rowh});
  \node[anchor=south east, inner sep=1pt, font=\small\itshape, text=stage3!90!black] at (-1.5,0.5*\rowh+0.04) {Lost after human review};
  \node[anchor=south west, inner sep=1pt, font=\small\itshape, text=stage1!90!black] at (1.5,0.5*\rowh+0.04) {Gained after human review};
\end{tikzpicture}
\caption{Bug families with the largest label changes during
independent review ($|\text{change}| \geq 3$). The remaining
23 families changed by fewer than 3 bugs or not at all (16
had no change). The reviewers changed the family of 64 retained bugs and removed two candidates. The largest shifts moved bugs out of Pass ordering
error ($-$17), Missing input validation ($-$11), and Stale
state error ($-$9), and into Compiler transformation error
($+$12). These four families show the largest net changes.}
\label{fig:review-changes}
\end{figure}

Cohen's $\kappa$ between the two authors' independent labels was 0.78 for bug family labels. The authors rejected two candidates, retaining 410 bugs. They revised the proposed labels for 64 bugs (15.6\%) and kept the other 346 (84.4\%). Of the 410 retained bugs, 391 have a single family and 19 have more than one (17 with two families and 2 with three), giving 431 family assignments across 36 observed families. Figure~\ref{fig:review-changes} summarizes the largest classification changes.

\subsection{Cross-Source Comparison}
\label{sec:comparison-method}

We recorded the presence of each bug family in mutation tools, curated benchmarks, and the verified repository corpus. Families can occur in any combination of these sources. A repository bug was considered represented by an earlier source if at least one of its assigned families occurred in the mutation-tool or curated-benchmark baseline; it was \emph{in-the-wild only} if none did. We report family counts separately from bug counts because a bug may belong to multiple families. Section~\ref{sec:results} presents the resulting overlap and representation gaps.

\section{Results}
\label{sec:results}

This section answers the three research questions using the 410 verified bugs described in Section~\ref{sec:classify-wild}. RQ1 (Section~\ref{sec:rq1}) reports the bug families that the mutation tools support. RQ2 (Section~\ref{sec:rq2}) reports the families in the curated real bug benchmarks. RQ3 (Section~\ref{sec:comparison}) compares both with the in-the-wild bugs. Throughout, we distinguish three units: a \emph{bug} is one verified issue, an
\emph{assignment} links one bug to one family, and a \emph{family} is counted once regardless of how many bugs it contains.

Table~\ref{tab:repository-summary} shows how the bugs distribute across the ten repositories.

\subsection{RQ1: Mutation-Supported Bug Families}
\label{sec:rq1}

The four mutation tools together support 20 bug families: 11
quantum-specific and 9 classical that QMutPy inherits from MutPy
(Table~\ref{tab:baseline-mutation}).
Table~\ref{tab:baseline-mutation} shows support for the 11
quantum-specific families. Most of this support is for gate
edits. \emph{Extra gate}, \emph{Missing gate}, and \emph{Wrong
gate} are the only families that all four tools support. \emph{Perturbed
parameter} appears in three tools, and each of the other seven
quantum-specific families appears in one or two. Four of them
rely on QuanForge alone and rest on source evidence only, because
we did not execute QuanForge (Section~\ref{sec:data-collection}). QuanForge's code also names an operator that sets a
parameter to zero, but the operator has no implementation, so we do
not count it. 

The nine classical families cover changes to values and logic
(\emph{Wrong arithmetic}, \emph{Wrong comparison}, \emph{Wrong
conditional}, \emph{Wrong constant}), to code structure
(\emph{Wrong decorator}, \emph{Wrong loop control},
\emph{Mishandled exception}, \emph{Broken inheritance}), and to
general statements (\emph{Missing statement}). They contain no
quantum-specific logic, but they still matter: quantum programs
and frameworks are largely Python code. Of the nine, only
\emph{Wrong constant} and \emph{Wrong decorator} appear among
the 410 in-the-wild bugs (Section~\ref{sec:comparison}).

None of the 20 families targets compiler passes, serialization,
dependency versions, or interactions between components.
Section~\ref{sec:comparison} shows that these are among the most
frequent bugs in the repository corpus.

\subsection{RQ2: Curated Real Bug Families}
\label{sec:rq2}

The curated sources contain 310 bugs and patterns, which group
into 35 families. All 20 mutation-supported families occur in the
curated inventory, which adds 15 families not represented by
mutation tools (Table~\ref{tab:curated-families}).

The most frequent curated families are \emph{Incorrect operator}
(48), \emph{Constant replacement} (42), and \emph{Conditional
error} (33), which together cover 39.7\% of the curated bugs.
Most curated cases are Qiskit programs that fail in how they
build a circuit or call the API. A curated benchmark keeps a
case when the bug can be isolated in a short program with a
known fix, which favors these program-level bugs.

The curated families are concentrated: 13 families account for
80.6\% of the bugs, and 8 families have a single case. Most
cases come from Qiskit, and most come from 2020 or earlier.
Apart from \emph{Zeroed parameter}, a gate-parameter fault that no
selected mutation tool implements, the other 14 families added
beyond mutation share one theme: they describe how a program uses
its framework. Four concern APIs and
interfaces (\emph{API issue}, \emph{Registration error},
\emph{Missing import}, \emph{Refactor}), two concern compilation
calls (\emph{Missing transpilation}, \emph{Missing
decomposition}), and three concern object state (\emph{Missing
method call}, \emph{Missing field set}, \emph{Confusing mutator
with accessor}). The remaining five involve values and control
logic, such as \emph{Conditional error} and \emph{Reversed
output convention}. Curated benchmarks thus add a layer that
mutation operators miss, recorded mainly for Qiskit and for the
platform versions of their collection period.

\subsection{RQ3: Comparison with In-the-Wild Bugs}
\label{sec:comparison}
Most in-the-wild bugs belong to families that neither earlier source contains. One likely explanation is the pace of change: quantum platforms change faster than the resources built to test them. The mutation tools and curated benchmarks reflect the platform features and software versions that existed when they were created, whereas the in-the-wild bugs were reported between July 2024 and July 2026. For example, Qiskit released versions 2.0 through 2.5 in this period, and releases 2.4 and 2.5 alone added more than 70 new features~\cite{qiskit_release_240,qiskit_release_250}. On the other hand, the curated benchmarks preserve bugs from earlier platform versions. Bugs4Q-NA~\cite{bugs4qna_repo} and Bug-Pattern~\cite{zhao2021bugpatterns,bugpattern_repo} collect earlier Qiskit bugs and bug patterns, and Bugs in Quantum Computing Platforms~\cite{paltenghi2022bugs,quantumplatformbugs_repo} collects bugs from 18 platforms up to its 2022 study. They therefore cannot contain bugs introduced by later compiler changes, new APIs, new input representations, or dependency updates.

The mutation tools overlap least with the in-the-wild bugs because they mutate the \emph{program under test}, whereas the in-the-wild bugs, which come from platform repositories, occur in the \emph{platform software} that constructs, transforms, compiles, serializes, or executes programs. Mutation operators mainly make local changes to circuits or to the Python code that constructs them, such as changing a gate, qubit, measurement, parameter, constant, or decorator. Such operators can produce an incorrect input program, but none of them targets a faulty compilation step, a wrong conversion between intermediate representations, an unhandled input type, or an incompatibility between library versions.

\begin{figure}[tbp]
\centering
\def\fsBarW{2.2}
\def\fsRow{0.28}
\def\fsBarH{0.17}
\def\fsLab{-3.45}
\def\fsLeft{-3.7}
\def\fsRight{2.85}
\colorlet{fsQ}{stage1}
\colorlet{fsM}{stage3!78}
\colorlet{fsC}{black!22}
\definecolor{fsSlate}{HTML}{5F6B7A}
\newcommand{\fspanel}[4]{%
  \pgfmathsetmacro{\fsa}{-#1*\fsRow+0.95*\fsRow}%
  \pgfmathsetmacro{\fsb}{-#2*\fsRow-0.9*\fsRow}%
  \filldraw[fill=white, draw=#3!35, line width=0.5pt, rounded corners=2.5pt] (\fsLeft,\fsa) rectangle (\fsRight,\fsb);
  \node[anchor=center, inner sep=0pt, font=\scriptsize\bfseries, text=#3!80!black] at ({0.5*(\fsLeft+\fsRight)},{-#1*\fsRow+0.05}) {#4};}
\newcommand{\fsrow}[6]{%
  \pgfmathsetmacro{\fsy}{-#1*\fsRow}%
  \pgfmathsetmacro{\fsq}{#4/#3*\fsBarW}%
  \pgfmathsetmacro{\fsm}{#5/#3*\fsBarW}%
  \pgfmathsetmacro{\fsc}{#6/#3*\fsBarW}%
  \pgfmathsetmacro{\fsqm}{\fsq+\fsm}%
  \node[anchor=east, inner sep=0pt, font=\scriptsize] at (-0.1,\fsy) {#2};
  \ifdim\fsq pt>0pt \fill[fsQ] (0,\fsy-0.5*\fsBarH) rectangle ++(\fsq,\fsBarH); \fi
  \ifdim\fsm pt>0pt \fill[fsM] (\fsq,\fsy-0.5*\fsBarH) rectangle ++(\fsm,\fsBarH); \fi
  \ifdim\fsc pt>0pt \fill[fsC] (\fsqm,\fsy-0.5*\fsBarH) rectangle ++(\fsc,\fsBarH); \fi
  \ifdim\fsq pt>0pt \ifdim\fsq pt<\fsBarW pt \draw[white, line width=0.6pt] (\fsq,\fsy-0.5*\fsBarH) -- ++(0,\fsBarH); \fi\fi
  \ifdim\fsm pt>0pt \ifdim\fsc pt>0pt \draw[white, line width=0.6pt] (\fsqm,\fsy-0.5*\fsBarH) -- ++(0,\fsBarH); \fi\fi
  \node[anchor=west, inner sep=0pt, font=\scriptsize, text=black!75] at (\fsBarW+0.1,\fsy) {#3};}
\newcommand{\fsaxis}[1]{%
  \pgfmathsetmacro{\fsy}{-#1*\fsRow}%
  \draw[black!45, line width=0.3pt] (0,\fsy+0.07) -- (\fsBarW,\fsy+0.07);
  \foreach \p/\t in {0/0, 0.5/50, 1/100\%} {
    \draw[black!45, line width=0.3pt] (\p*\fsBarW,\fsy+0.07) -- ++(0,-0.05);
    \node[anchor=north, inner sep=1pt, font=\tiny, text=black!60] at (\p*\fsBarW,\fsy+0.02) {\t};}}
\newcommand{\fskey}[3]{%
  \fill[#3] (#1,-#2*\fsRow-0.5*\fsBarH) rectangle ++(0.3,\fsBarH);}
\begin{tikzpicture}[x=1cm, y=1cm]
\begin{scope}
\fspanel{0}{8}{stage1}{In all three sources (8 families, 23 bugs)}
\fsrow{1}{Wrong qubit target}{10}{8}{2}{0}
\fsrow{2}{Wrong gate}{4}{4}{0}{0}
\fsrow{3}{Extra gate}{3}{2}{1}{0}
\fsrow{4}{Missing gate}{2}{2}{0}{0}
\fsrow{5}{Flipped parameter sign}{1}{1}{0}{0}
\fsrow{6}{Misplaced gate}{1}{1}{0}{0}
\fsrow{7}{Wrong constant}{1}{0}{1}{0}
\fsrow{8}{Wrong decorator}{1}{0}{1}{0}
\fspanel{10.3}{18.3}{stage2}{Curated and in-the-wild (7 families, 34 bugs)}
\fsrow{11.3}{Conditional error}{12}{7}{5}{0}
\fsrow{12.3}{Missing field set}{9}{4}{4}{1}
\fsrow{13.3}{API issue}{5}{4}{1}{0}
\fsrow{14.3}{Incorrect operator}{3}{2}{1}{0}
\fsrow{15.3}{Incorrect parameters}{2}{1}{1}{0}
\fsrow{16.3}{Missing method call}{2}{1}{1}{0}
\fsrow{17.3}{Constant replacement}{1}{0}{1}{0}
\fsaxis{18.3}
\filldraw[fill=white, draw=black!25, line width=0.5pt, rounded corners=2.5pt] (\fsLeft,{-19.75*\fsRow}) rectangle (\fsRight,{-22.9*\fsRow});
\fskey{-3.4}{20.65}{fsQ}\node[anchor=west, inner sep=0pt, font=\scriptsize] at (-3.0,-20.65*\fsRow) {Quantum-specific};
\fskey{-0.45}{20.65}{fsM}\node[anchor=west, inner sep=0pt, font=\scriptsize] at (-0.05,-20.65*\fsRow) {Quantum--classical};
\fskey{-3.4}{21.95}{fsC}\node[anchor=west, inner sep=0pt, font=\scriptsize] at (-3.0,-21.95*\fsRow) {Purely classical};
\node[anchor=west, inner sep=0pt, font=\scriptsize\itshape, text=black!60] at (-0.45,-21.95*\fsRow) {$n$ = bugs in the family};
\end{scope}
\begin{scope}[shift={(6.85,0)}]
\fspanel{0}{22}{fsSlate}{In-the-wild only (21 families, 361 bugs)}
\fsrow{1}{Compiler transformation error}{66}{49}{10}{7}
\fsrow{2}{Type-handling error}{53}{32}{15}{6}
\fsrow{3}{Missing input validation}{47}{31}{8}{8}
\fsrow{4}{Circuit construction error}{29}{15}{4}{10}
\fsrow{5}{Dependency incompatibility}{28}{11}{8}{9}
\fsrow{6}{Missing compilation step}{23}{15}{5}{3}
\fsrow{7}{Format conversion error}{20}{13}{3}{4}
\fsrow{8}{Serialization error}{18}{10}{6}{2}
\fsrow{9}{Visualization error}{18}{13}{2}{3}
\fsrow{10}{Mathematical formulation error}{13}{11}{1}{1}
\fsrow{11}{Numerical error}{11}{5}{3}{3}
\fsrow{12}{Pass ordering error}{11}{9}{1}{1}
\fsrow{13}{Measurement ordering error}{6}{5}{0}{1}
\fsrow{14}{Stale state error}{6}{5}{0}{1}
\fsrow{15}{Equality semantics error}{5}{4}{0}{1}
\fsrow{16}{Measurement result error}{5}{4}{0}{1}
\fsrow{17}{Copy aliasing error}{4}{2}{2}{0}
\fsrow{18}{Symbolic algebra error}{4}{1}{3}{0}
\fsrow{19}{Backend dispatch error}{3}{3}{0}{0}
\fsrow{20}{Classical control error}{2}{2}{0}{0}
\fsrow{21}{Concurrency error}{2}{2}{0}{0}
\fsaxis{22}
\end{scope}
\end{tikzpicture}
\caption{The 36 bug families observed in the 410 in-the-wild bugs, grouped by the sources that contain them. Each bar is the family's bugs scaled to 100\% and split into quantum-specific, quantum--classical, and purely classical bugs (Section~\ref{sec:classify-wild}). $n$ counts the bugs assigned to the family, so a bug with several families is counted in each (431 assignments for 410 bugs). The in-the-wild-only panel counts the 361 bugs with at least one such family. Seven of them also have a baseline family, so 354 bugs have only in-the-wild-only families. No family is mutation-supported only, because every mutation-supported family also occurs in the curated benchmarks. Family definitions are in the \href{https://figshare.com/s/cadac005ec0928cacb3b}{replication package}.}
\label{fig:family-specificity}
\end{figure}
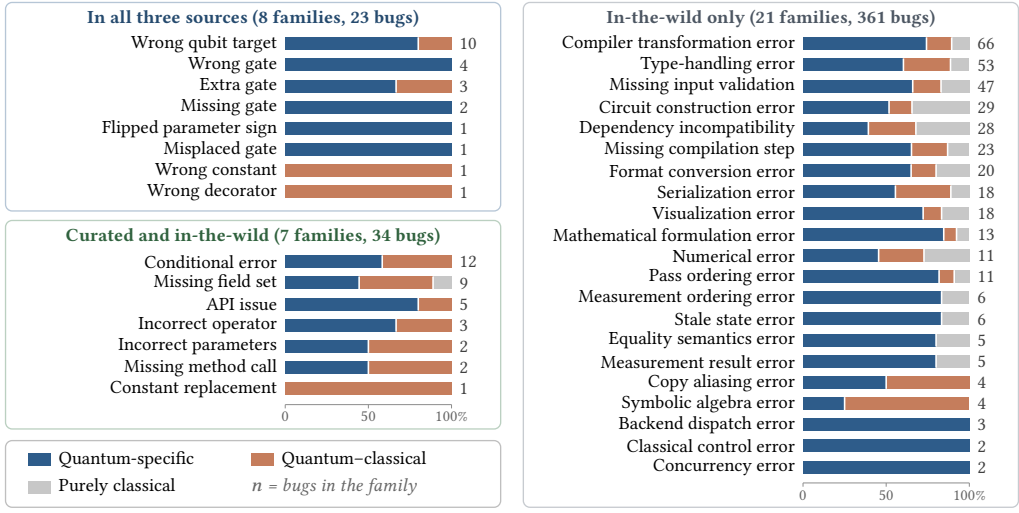

Only 23 of the 410 bugs (5.6\%) have a mutation-supported family, and 56 (13.7\%) have a family from either earlier source. The other 354 (86.3\%) belong only to families that neither source contains, and they concentrate in compilers, data formats, input handling, and dependencies. The gap does not come from the two largest repositories. Without Qiskit and PennyLane, 89 of the remaining 100 bugs (89.0\%) are in-the-wild only, and the share is between 80.7\% and 95.5\% in each of the five largest repositories (Table~\ref{tab:repository-summary}). At the family level, 15 of the 36 observed families appear in the 35-family baseline, and 21 are absent from both earlier sources (Figure~\ref{fig:family-specificity}). Of the 56 families across the three sources, the other 20 have no in-the-wild bug: 12 that the mutation tools and curated benchmarks share, and 8 that only the curated benchmarks contain.  

\vspace{.3em}
\noindent\textbf{Mutation tools match in-the-wild bugs only on
local edits.}
The eight families shared by mutation tools and in-the-wild bugs are
\emph{Wrong qubit target}, \emph{Wrong gate}, \emph{Extra gate},
\emph{Missing gate}, \emph{Flipped parameter sign},
\emph{Misplaced gate}, \emph{Wrong constant}, and \emph{Wrong
decorator}. They cover 23 bugs, led by \emph{Wrong qubit target}
(10) and \emph{Wrong gate} (4). Each is a change at a single
location.
Twelve of the 20 mutation-supported families do not appear among
the 410 bugs, including both measurement families, two of the
three parameter families, and seven of the nine classical
families. Much of the tools' bug model therefore targets bugs
that we did not observe in the ten repositories.

\vspace{.3em}
\noindent\textbf{Curated benchmarks add bugs in how code uses
the framework.}
Seven curated families without mutation support appear in 34
bugs, led by \emph{Conditional error} (12), \emph{Missing field
set} (9), and \emph{API issue} (5). These bugs sit in API
contracts, object fields, method calls, and control logic rather
than in the circuit. Curated benchmarks thus cover a layer that
the mutation operators miss, but this layer accounts for only 34
of the 431 assignments (7.9\%).

\vspace{.3em}
\noindent\textbf{In-the-wild-only bugs depend on inputs, compiler
passes, and library versions.}
The 21 in-the-wild-only families cover 361 bugs and 374
assignments (Figure~\ref{fig:family-specificity}). Five of them account
for 51.7\% of all assignments: \emph{Compiler transformation
error} (66), \emph{Type-handling error} (53), \emph{Missing
input validation} (47), \emph{Circuit construction error} (29),
and \emph{Dependency incompatibility} (28). These bugs typically
appear only under specific conditions. A compiler transformation
error appears when a specific sequence of passes rewrites a
circuit. A type-handling error appears when a valid but
unexpected input type or shape reaches the code. A dependency
incompatibility appears with a specific library version.
Serialization and stale-state errors appear when state is saved
and reloaded, or reused after a change. Neither a local mutation
of the program nor a benchmark collected from earlier releases
recreates these conditions.

\vspace{.3em}
\noindent\textbf{Quantum-specific bugs dominate most families, and
purely classical bugs dominate none.}
Of the 410 bugs, 271 (66.1\%) are quantum-specific, 82 (20.0\%)
are quantum--classical, and 57 (13.9\%) are purely classical
(Section~\ref{sec:classify-wild}). Figure~\ref{fig:family-specificity}
shows this split for each family. Quantum-specific bugs are the
majority in 26 of the 36 observed families, for example
\emph{Compiler transformation error} (49 of 66), \emph{Wrong qubit
target} (8 of 10), and \emph{Conditional error} (7 of 12).
Quantum--classical bugs are the majority in four families:
\emph{Symbolic algebra error} (3 of 4) and the single-bug families
\emph{Wrong constant}, \emph{Wrong decorator}, and \emph{Constant
replacement}. The other six families have no majority, among them
\emph{Dependency incompatibility} (11 quantum-specific, 8
quantum--classical, and 9 purely classical bugs) and \emph{Missing
field set} (4, 4, and 1). In no family are purely classical bugs the majority.
The 57 purely classical bugs spread over 17 families and are most common
in \emph{Circuit construction error} (10), \emph{Dependency
incompatibility} (9), and \emph{Missing input validation} (8).
Quantum--classical bugs outnumber purely classical ones because of
PennyLane, which couples circuits to PyTorch, JAX, and GPU back
ends: 42 of the 82 are PennyLane bugs, and 31\% of PennyLane's
bugs are quantum--classical against 10\% of Qiskit's. Nor is the gap
between the sources a matter of classical code that quantum
testing tools could leave aside. Of the 354 in-the-wild-only bugs,
234 (66.1\%) are
quantum-specific, the same share as in the whole corpus.

\vspace{.3em}
\noindent\textbf{Many in-the-wild bugs are in fast-changing parts
of the quantum platforms.}
Compilers, which the platforms keep extending, show this most
clearly. TKET and CUDA-Q, the two projects built around a
compiler, have no mutation-supported bugs, and most of their bugs
are in-the-wild only: 21 of 22 in TKET and 19 of 23 in CUDA-Q
(Table~\ref{tab:repository-summary}). Qiskit issue
\#16594~\cite{qiskit16594issue} shows how such bugs arise. In Qiskit 2.5.0, the
\texttt{CommutativeCancellation} pass, which runs at
optimization levels 2 and 3, removed two \texttt{sxdg} gates whose product is X
when \texttt{sx} was a supported gate, so the compiled circuit
produced different measurement results. The same circuits
compiled correctly with Qiskit 2.4.0. The bug entered between
these two releases, in the pass's Rust implementation, which the
selected mutation tools cannot mutate because they operate on
circuits and Python programs.

\begin{table}[htbp]
\centering
\caption{Verified bugs in each repository and how they relate
to the earlier sources. Each bug is counted once. Bugs with a
mutation-supported family are counted first. Of the remaining
bugs, those with a family in the curated sources are counted
next. All others are counted as in-the-wild only.}
\label{tab:repository-summary}
\footnotesize
\setlength{\tabcolsep}{6pt}
\renewcommand{\arraystretch}{1.1}
\begin{tabular}{@{}l r r r r@{}}
\toprule
\textbf{Repository} & \textbf{Bugs}
  & \shortstack[r]{\textbf{Mutation-}\\\textbf{supported}}
  & \shortstack[r]{\textbf{Curated-}\\\textbf{supported}}
  & \shortstack[r]{\textbf{In-the-wild}\\\textbf{only}} \\
\midrule
Qiskit            & 175 & 8  & 11 & 156 \\
PennyLane         & 135 & 13 & 13 & 109 \\
Cirq              & 37  & 2  & 2  & 33  \\
CUDA-Q            & 23  & 0  & 4  & 19  \\
TKET              & 22  & 0  & 1  & 21  \\
QuTiP             & 5   & 0  & 1  & 4   \\
Amazon Braket SDK & 4   & 0  & 0  & 4   \\
Mitiq             & 4   & 0  & 1  & 3   \\
OpenQASM          & 3   & 0  & 0  & 3   \\
PyQuil            & 2   & 0  & 0  & 2   \\
\midrule
\textbf{Total} & \textbf{410} & \textbf{23} & \textbf{33} & \textbf{354} \\
\bottomrule
\end{tabular}
\end{table}

\vspace{.3em}
\noindent\textbf{Some bugs belong to more than one family.}
Of the 410 bugs, 19 have more than one family (17 with two families
and 2 with three). In CUDA-Q issue \#4401~\cite{cudaq4401issue}, the compiler lowers a loop
comparison incorrectly and drops the final iteration, which combines \emph{Classical control error}, a wrong value in the classical logic that shapes the quantum program, with \emph{Compiler transformation error}. Reproducing such a bug requires both
mechanisms at once. Table~\ref{tab:wild-bugs} shows three
verified examples.

The earlier sources thus represent only a small part of the
in-the-wild bugs. Mutation tools match them only on local edits,
and curated benchmarks add bugs in how code uses the framework,
but most in-the-wild bugs belong to families that neither source
contains. These bugs concentrate in compilers, data formats, input
handling, and dependencies, so testing current quantum platforms
needs fault models that cover these parts as well.

\begin{table}[htbp]
\centering
\captionsetup{skip=4pt}
\caption{Examples of verified in-the-wild bug classifications.}
\label{tab:wild-bugs}
\footnotesize
\renewcommand{\arraystretch}{1.0}
\begin{tabularx}{\textwidth}{@{}
    >{\raggedright\arraybackslash}p{1.5cm}
    >{\raggedright\arraybackslash}p{1.5cm}
    >{\raggedright\arraybackslash}X
    >{\raggedright\arraybackslash}p{3.1cm}@{}}
\toprule
\textbf{Repository} & \textbf{Issue}
  & \textbf{Observed defect}
  & \textbf{Verified families} \\
\midrule
Qiskit
  & \#16594~\cite{qiskit16594issue}
  & An optimization pass removes two \texttt{sxdg} gates whose
    product is X and changes circuit semantics.
  & Compiler transformation error \\
CUDA-Q
  & \#4401~\cite{cudaq4401issue}
  & Compiler lowering changes a loop comparison and omits the
    final iteration.
  & Classical control error; Compiler transformation error \\
Qiskit
  & \#15613~\cite{qiskit15613issue}
  & QPY cannot reload a circuit containing a register whose name
    begins with \texttt{ancilla}.
  & Serialization error \\
\bottomrule
\end{tabularx}
\end{table}

\subsection{Threats to Validity}
\label{sec:threats}

\subsubsection{Internal and Construct Validity}
\leavevmode\\
\textbf{Coding subjectivity.} Assigning a bug to a family requires interpreting its cause, observed behavior, and likely fix, particularly when an issue could fit multiple categories. Two of the authors independently verified every AI classification and could add, remove, rename, or otherwise revise families; conflicts were resolved through discussion and consensus. Review changed 64 retained classifications and removed two candidates, demonstrating that the human stage materially affected the result. The reported 84.4\% agreement is agreement with the supplied proposal and may be inflated because both reviewing authors saw that proposal. Multiple names are preserved when distinct interacting mechanisms apply.

\noindent \textbf{Automated screening notes.} Many screening notes and fault descriptions came from templates and keyword rules applied in batch, so they are less specific than per-issue judgments. They served only to propose families, and the two authors' review of every retained bug is the check on them.

\noindent \textbf{Incomplete evidence.} GitHub issue descriptions and comments do not always include the final patch, regression test, or complete root-cause analysis. The model received truncated excerpts, although the authors used the available repository evidence when verifying eligibility and family names. Reports with insufficient evidence could be excluded or classified at a broader mechanism level, and the assigned family still reflects only the evidence available in the record.

\noindent \textbf{Families derived from the corpus.} Five family definitions came from a random sample of repository issues, so they describe this corpus by construction. A different sample could name and bound them differently.

\subsubsection{External Validity}
\leavevmode\\
\textbf{Selection and temporal bias.} Our results depend on the four mutation tools, three curated datasets, and ten GitHub repositories selected for the study. The in-the-wild corpus was collected on August 22 and August 25, 2026, and covers issues created from July 24, 2024 through July 23, 2026. It was retrieved through repository-specific bug issue types or labels without free-text bug keywords. The snapshot therefore excludes later reports and can miss unlabeled or differently labeled defects, bugs from other repositories, and reports omitted during earlier screening. Of the 1,019 collected issues, 607 never reached author review: 413 were Tier~3 issues that maintainers had not accepted, 77 were not confirmed as bugs by the agent, 110 lacked a verified fault description, and 7 were excluded or held afterwards. Some real bugs in these groups may therefore be missing from the corpus, so the counts for some families may be too low. Qiskit and PennyLane contribute 310 of 410 bugs, so their development practices strongly influence the distribution. Requiring concrete evidence improves confidence in retained reports but can exclude genuine defects whose issue records lack detailed reproductions, traces, or expected-versus-observed behavior.

\noindent \textbf{Generalizability.} The observed distribution may not generalize to all quantum projects, programming languages, libraries, hardware providers, or development practices. Other repositories, periods, or releases may exhibit different family distributions. The reported shares describe bugs resembling those from these ten repositories and this period. They do not support claims about all quantum software. Recording name-based source membership enables comparison but still simplifies differences among individual contexts. The results should therefore be interpreted as evidence about the selected corpus rather than as an exhaustive taxonomy.

\section{Discussion and Lessons}

Our findings point to a distinction that matters for quantum
programming language research: testing a quantum program is
not the same as testing the system that processes it.
Existing mutation tools primarily introduce faults into
programs and circuits, while many recent bugs arise in the
compilers, representations, and framework interfaces that
interpret and transform them. Only 23 of the 410 bugs
(5.6\%) belong to mutation-supported families, while
354 (86.3\%) belong to families absent from both mutation
tools and curated benchmarks. These numbers measure fault
representation, not the detection capability of individual
testing techniques. We draw four lessons from this gap.

\vspace{.3em}
\noindent\textbf{Lesson 1: Correct circuits do not guarantee
correct compilation.}
Compiler transformation errors form the largest uncovered
family, with 66 bugs (Figure~\ref{fig:family-specificity}).
These faults arise when compilation changes the behavior of an
otherwise valid program. In Qiskit issue
\#16594~\cite{qiskit16594issue}, an optimization pass removes
two gates that do not cancel, so the compiled circuit gives
different measurement results (Section~\ref{sec:comparison},
Table~\ref{tab:wild-bugs}). The input circuit is correct, and
the failure appears only at optimization levels 2 and 3, where
this pass runs. A mutation operator that modifies the input
circuit therefore cannot reproduce this fault. This suggests a different
testing target: the transformations performed by the compiler
itself. Tests should check whether individual passes and their
compositions preserve observable program behavior across
optimization settings and target configurations. For the
programming languages community, the lesson is that
semantic preservation must be tested at the compilation
boundary, not assumed from the correctness of the input
program.

\vspace{.3em}
\noindent\textbf{Lesson 2: Representation boundaries deserve
explicit correctness checks.}
Many uncovered faults concern how programs and their data
are represented and exchanged. Figure~\ref{fig:family-specificity}
shows 53 type-handling errors, 47 cases of missing input
validation, 20 format conversion errors, and 18 serialization
errors. In Qiskit issue \#15613~\cite{qiskit15613issue}, QPY,
Qiskit's format for saving circuits, cannot reload a circuit
that contains a register whose name begins with
\texttt{ancilla} (Table~\ref{tab:wild-bugs}). The circuit's
gates are not the problem. The failure occurs when the saved
circuit is read back. These
findings motivate testing the contracts between
representations, including accepted input types,
intermediate-representation invariants, conversion
correctness, and serialization round trips. Such properties
are relevant beyond quantum software, wherever language
implementations translate programs between representations.

\vspace{.3em}
\noindent\textbf{Lesson 3: Fault models must capture context,
not just local edits.}
Several frequent uncovered families require conditions
that a local mutation cannot express. Compiler bugs depend on
the transformation pipeline: Qiskit issue \#16594 fails only
at the optimization levels that run the faulty pass.
Dependency incompatibilities (28 bugs,
Figure~\ref{fig:family-specificity}) depend on library
versions: in PennyLane issue \#9185, ZX circuit transformations
fail with PyZX 0.10.0~\cite{pennylane9185issue}. Other bugs
combine several mechanisms, and 19 of the 410 bugs have more
than one family (Section~\ref{sec:comparison}). In CUDA-Q issue
\#4401~\cite{cudaq4401issue}, the compiler lowers a greater-than-or-equal loop condition as strictly greater than, so the loop skips its final iteration. The fault is in the compiler, not in the user's program. This one bug is both a
\emph{Classical control error} and a \emph{Compiler
transformation error} (Table~\ref{tab:wild-bugs}). These cases call for fault models that
operate at different levels of the programming system.
Some could be represented by mutating compiler rewrite
rules, pass behavior, or representation-handling logic.
Others require preserving a particular input, configuration,
or dependency environment. The goal should not simply be
more gate-level operators, but fault models that reproduce
the conditions under which real failures occur.

\vspace{.3em}
\noindent\textbf{Lesson 4: Benchmarks must preserve failure
conditions and evolve with the software.}
21 bug families observed in recent repositories are
absent from both earlier sources. Existing curated benchmarks
preserve valuable real faults, but their coverage reflects
the projects and versions from which they were collected.
Adding a bug report alone is insufficient when reproducing
the failure requires a particular compiler pipeline, input
representation, or dependency version. Future benchmarks
should retain these conditions alongside the buggy version,
fix, and regression test, and incorporate newly observed
fault mechanisms as platforms evolve.

This also has a methodological implication. Nineteen bugs
in our corpus involve multiple families, and independent
review changed 64 proposed classifications. A useful fault
benchmark therefore needs an auditable mapping from each
bug to its underlying mechanisms, with bug counts kept
distinct from family assignments. Otherwise, different
classification decisions can change the apparent coverage
of a testing approach.

\noindent\textbf{Overall lesson.}
Quantum software testing must look beyond faults in
quantum circuits to faults in the programming systems
that give those circuits their executable meaning.
Our results identify compiler transformations,
representation boundaries, and execution context as
important targets for future testing techniques,
fault models, and benchmarks.

\section{Conclusion}

We investigated how well existing quantum mutation tools and
curated bug benchmarks represent faults found in modern quantum
software. By comparing four mutation tools, three curated
benchmarks, and 410 verified bugs from ten repositories under
a common taxonomy of 56 bug families, we found a substantial
representation gap. Only 23 bugs (5.6\%) belong to families
supported by mutation tools, while 354 (86.3\%) belong to
families absent from both mutation tools and curated benchmarks.
Twenty-one bug families observed in recent repositories are
missing from these earlier sources.

More importantly, the gap reveals where existing fault models
fall short. While mutation tools primarily model local changes
to circuits and programs, many real faults arise in compiler
transformations, type handling, input validation, and
interactions across software components. These faults often
depend on program representations, compilation pipelines, or
execution context that isolated mutations cannot reproduce.
Our findings suggest that testing quantum programs is not
enough; we must also test the programming systems that
transform and execute them. Future fault models and benchmarks
should therefore account for compiler semantics, representation
boundaries, and the conditions needed to reproduce real faults.

\section*{Data-Availability Statement}
The dataset for this study is available to reviewers at
\url{https://figshare.com/s/cadac005ec0928cacb3b}. 


\bibliographystyle{ACM-Reference-Format}
\bibliography{main}
\end{document}